\documentclass[letterpaper]{article} 
\usepackage{aaai2027}  
\usepackage[hyphens]{url}  
\usepackage{graphicx} 
\usepackage{natbib}  
\usepackage{caption} 
\usepackage{algorithm}
\usepackage{algorithmic}

\usepackage{newfloat}
\usepackage{listings}
\DeclareCaptionStyle{ruled}{labelfont=normalfont,labelsep=colon,strut=off} 
\floatstyle{ruled}
\newfloat{listing}{tb}{lst}{}
\floatname{listing}{Listing}

\usepackage{booktabs}

\title{What Do Sound Event Embeddings Reveal? Auditing Speaker Leakage in Audio Representations}
\author{
   Bhawana Chhaglani
}
\affiliations{
    Penn State University\\
    bbc5655@psu.edu

}

\begin{document}

\maketitle

\begin{abstract}
Pretrained audio embeddings are increasingly used as intermediate
representations for sound event detection (SED) and other always-on
acoustic sensing tasks. While these representations are selected for
their utility on downstream tasks, they may also retain sensitive
information unrelated to their intended purpose, creating privacy risks
even when raw audio is not exposed. 
In this work, we present an empirical framework for auditing privacy
leakage from pretrained audio embeddings using two complementary
approaches. First, we quantify representation-level leakage by measuring
statistical dependence between an audio embedding and a representation
specialized for encoding sensitive information. Second, we measure
attack-level leakage by training lightweight adversarial probes to
recover sensitive attributes directly from the embedding. We apply this
framework to four widely used audio representations (CLAP, PANNs, AST,
and YAMNet) and evaluate leakage of speaker identity and gender.
We found that speaker identification balanced accuracy ranges from $0.43$ to $0.72$, while gender information is also strongly recoverable. Moreover,
representation-level and attack-based measurements exhibit consistent
trends, with CLAP leaking the most and YAMNet leaking the lest speaker identity. 
These results demonstrate that intermediate audio representations are
not inherently privacy-preserving interfaces and motivate privacy
leakage as an explicit criterion, alongside downstream utility, when
selecting representations for acoustic sensing systems.
\end{abstract}

\section{Introduction}

\begin{figure*}
    \centering
    \includegraphics[width=0.8\linewidth]{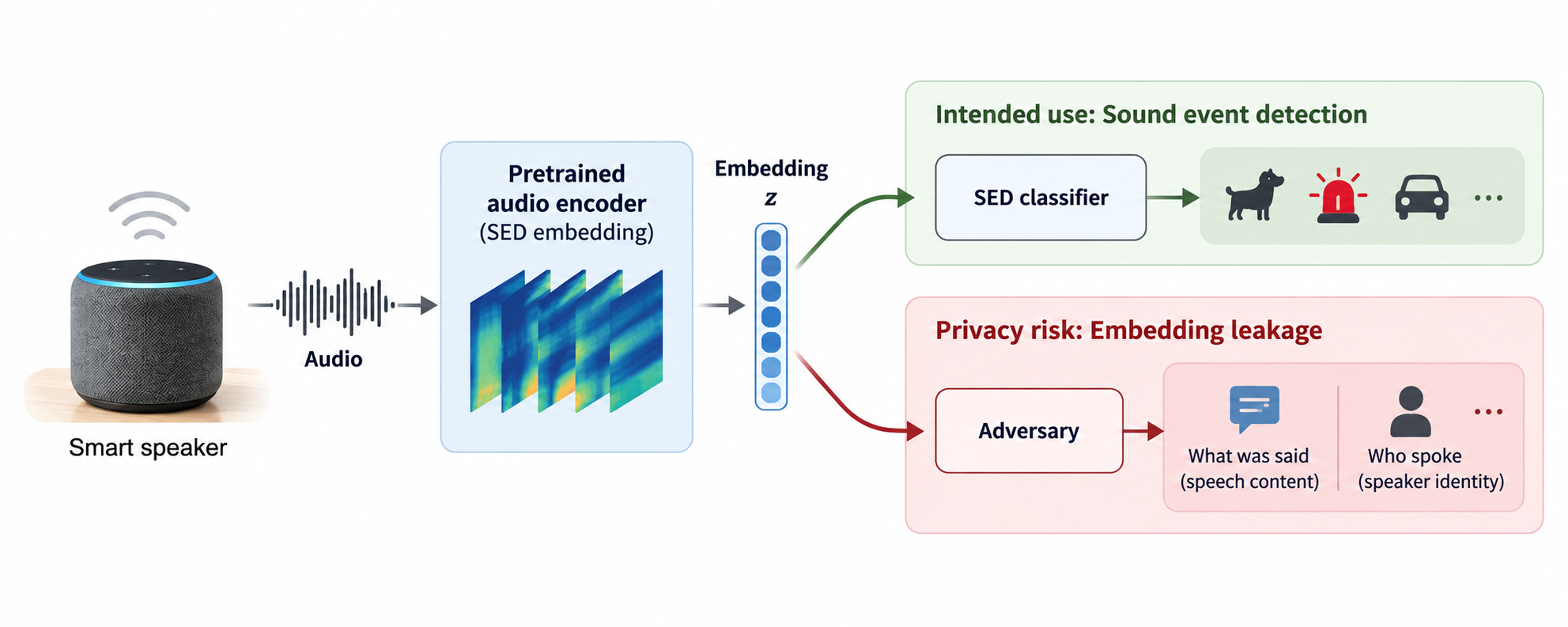}
    \caption{\small Audio embeddings used for SED on smart speakers}
    \label{fig:intro}
    \vspace{-0.6cm}
\end{figure*}

Audio sensing is increasingly used in smart speakers, wearables, and
other always-on devices to recognize environmental sounds such as
alarms, glass breaking, coughing, and other acoustic events \cite{amazon2024alexaemergencyassist,heittola2013context, chhaglani2024cough}. Rather than
operating directly on raw audio, modern sound event detection (SED)
systems often rely on pretrained audio encoders that transform an input
signal $X$ into an intermediate representation $Z$, which is then sent to cloud and used
for a downstream task $T$ as shown in Figure \ref{fig:intro}. Models such as CLAP \cite{elizalde2023clap}, PANNs \cite{kong2020panns}, AST \cite{gong2021ast}, and YAMNet \cite{korkmaz2025performance}
provide general-purpose representations that make it easier to build
such systems without training an audio encoder from scratch. They have been shown to be effective for various applications. For example, YAMNet has been used for cardiovascular disease classification \cite{barnawi2023evaluation}, speech emotion recognition \cite{boateng2020speech}, COVID detection \cite{sobahi2022explainable}, and identifying parkinsonian speech \cite{syed2020deep}. Similarly, CLAP has been used for heart rate estimation and auscultation \cite{nie2025foundation}, and SED \cite{elizalde2023clap} (including emergency sounds).

However,
although these embeddings are intended to capture information useful
for acoustic understanding, they are computed from the entire acoustic
signal and may also encode information unrelated to the intended tasks.
When speech occurs in the environment, this can include sensitive
information about the people being recorded, even when the raw audio
itself is never exposed \cite{williams2023new,chhaglani2024towards,chen2008audio,boovaraghavan2024kirigami}.
This raises a fundamental question for privacy-aware audio sensing:
\emph{how do we know whether an audio embedding is leaking private
information?} 

Ideally, a representation should retain information about
its intended task(s) $T$ while discarding information about a sensitive
attribute(s) $S$, i.e., $I(Z;S) \ll I(Z;T)$. Yet audio embeddings are
typically selected based on downstream utility, with considerably less
attention to what unintended information remains encoded in $Z$.
Consequently, two embeddings that provide similar functionality may
expose very different amounts of information about the user. This is
particularly important for intermediate representations that are stored,
transmitted, or made accessible to downstream applications under the
assumption that they are safer than raw audio.

In this work, we investigate this question by auditing \emph{speaker
privacy} in commonly used audio embeddings. We study four commonly used pretrained audio
representations, CLAP, PANNs, AST, and YAMNet - using two complementary
measures of leakage. First, we introduce a \emph{representation-level}
measure that asks how much information in an audio embedding overlaps
with a representation explicitly designed to encode speaker identity.
We extract ECAPA-TDNN \cite{guo2024ecapa} speaker embeddings from the same audio and measure
statistical dependence between the two representation spaces using
canonical correlation analysis (CCA) and mutual-information (MI) estimation.
Second, we perform an \emph{attack-level} evaluation in which an
adversary receives only the audio embedding and trains a lightweight
classifier to infer sensitive attributes. We consider speaker
identification and gender inference, providing an operational measure
of whether information retained in an embedding can actually be
exploited.

We evaluate these approaches using the
Common Voice dataset spanning five speakers (three male and two female).
Our results reveal substantial and consistent differences across audio
embeddings. For speaker identification with only 2 seconds of enrollment
audio per speaker, balanced accuracy ranges from $0.43$ for YAMNet to
$0.72$ for CLAP. Our representation-level analysis shows the same
overall pattern: CLAP exhibits the greatest dependence on
speaker-specialized representations, while YAMNet exhibits the least.
Gender information is also recoverable across the evaluated
representations, with CLAP exhibiting the strongest leakage. 

Our findings highlight an overlooked dimension in selecting audio
representations for privacy-sensitive sensing systems. Intermediate
embeddings should not be assumed to be privacy-preserving simply because
raw audio is discarded: representations optimized for acoustic utility
can retain substantial information about the people present in the
signal. More broadly, our results motivate evaluating audio embeddings
along two dimensions: \emph{utility for the intended task} and
\emph{leakage of sensitive information} and treating privacy as an
explicit model-selection criterion. Such auditing can help developers
identify representations that preserve the functionality required by
an application while minimizing unnecessary exposure of user
information.

\section{Background}

Privacy in audio sensing can be examined at different
levels of the processing pipeline: signal-level,
feature-level, and model-level privacy.
\textit{Signal-level privacy.}
Signal-level approaches aim to suppress sensitive
information directly in the audio signal while
preserving information useful for acoustic sensing. This includes technique like low-pass filtering \cite{pohlhausen2025revisiting,iravantchi2021privacymic}, subsampling \cite{mollyn2022samosa}, removing likely speech segments \cite{boovaraghavan2024kirigami}, etc.
For example, Kirigami~\cite{boovaraghavan2024kirigami}
investigates speech-content leakage from signal-level privacy techniques. 
\textit{Feature-level privacy.}
Feature-level approaches carefully extract features extracted from audio signals relevant to the task. For example,  Symdetector \cite{sun2015symdetector} extract handcrafted features from audio to detect cough and sneeze. 
FeatureSense~\cite{chhaglani2025featuresense} evaluates privacy (speaker) leakage 
and
investigates the privacy-utility tradeoffs
across low-level handcrafted acoustic features, demonstrating that
features useful for sound classification may also
retain sensitive information. Such analysis helps
identify feature representations that balance
downstream sensing utility with privacy.
\textit{Model-level privacy.}
Model-level privacy techniques aim to ensure privacy at the ML model and prevent
models from learning or exposing
sensitive information while preserving their
utility for intended tasks. Common approaches
include differential privacy \cite{paul2025secure}, adversarial learning,
and privacy-preserving representation learning \cite{gharib2024adversarial}.
In contrast to methods that modify model training,
our work audits the privacy of intermediate
embeddings generated by existing pretrained audio
encoders. Specifically, we evaluate whether these
representations retain speaker identity and gender
information that can be recovered by an adversary.
Our work extends privacy
auditing from raw audio and handcrafted features
to intermediate neural representations.

\noindent\textbf{Challenges in Privacy Evaluation.}
Evaluating privacy leakage remains challenging because
sensitive information may persist in audio representations
even when it is not directly observable. Consequently,
subjective listening tests alone are insufficient, as
information unintelligible to humans may still be
recoverable by machine-learning models
~\cite{pohlhausen2025revisiting,boovaraghavan2024kirigami}. Privacy risks also
depend on the adversary's capabilities, available training
data, and the sensitive attributes being targeted (like speech, speaker or context). Moreover,
statistical dependence does not necessarily indicate whether
the information can be exploited in practice. These challenges
motivate complementary evaluation approaches that measure
both the information retained in a representation and its
recoverability through adversarial attacks.
Accordingly, our threat model assumes an adversary
who has access to pretrained audio embeddings, but
not the original audio, and attempts to infer
speaker identity and attributes using machine-learning probes.
\section{Auditing Privacy Leakage in Audio Embeddings}

\subsection{Problem Formulation}

Consider an audio encoder $f$ that maps an input audio signal $X$ to
an intermediate representation $Z=f(X)$. The representation is exposed
to a downstream application performing an intended task $T$, such as
sound event detection. However, $Z$ may simultaneously retain
information about sensitive attributes $S$ that are unnecessary for
the intended task, such as speaker identity or gender.
Conceptually, a privacy-preserving representation should maximize
information relevant to the intended task while minimizing information
about sensitive attributes:
    $I(Z;S) \ll I(Z;T)$.
We therefore consider an embedding to exhibit privacy leakage when
sensitive information remains encoded in $Z$ and can be detected or
recovered despite being unrelated to the intended application. We
evaluate such leakage from two complementary perspectives:
representation-level dependence and attack-level recoverability.

\subsection{Representation-Level Leakage}

Our first approach measures whether an audio representation contains
information associated with a sensitive attribute without explicitly
training an attacker to predict that attribute. We use a pretrained
ECAPA-TDNN speaker encoder to obtain a reference representation
$Z_S$ that is explicitly optimized for speaker discrimination. Given
an audio embedding $Z$ and the corresponding speaker embedding $Z_S$
for the same utterance, we measure dependence between the two
representation spaces.

Because the representations differ in dimensionality, we first
standardize them and apply PCA, followed by canonical correlation
analysis (CCA) to identify maximally correlated components. We then
estimate mutual information between paired canonical components using
a $k$-nearest-neighbor estimator. Higher dependence indicates that $Z$
retains more information aligned with $Z_S$.
We use this as a representation-level indicator of
speaker leakage rather than as a direct estimate of $I(Z;S)$.

\subsection{Attack-Level Leakage}

Our second approach asks whether sensitive information encoded in $Z$
can be exploited by an adversary. We assume an adversary has access to
the intermediate embedding but not the original audio and trains a
lightweight classifier $g:Z\rightarrow S$ to recover a sensitive
attribute. 
From an information-theoretic perspective, sensitive-information leakage
can be expressed as $I(Z;S)=H(S)-H(S|Z)$: the more observing $Z$ reduces
uncertainty about $S$, the more sensitive information the representation
reveals. Our attack-level evaluation operationalizes this notion by
measuring how accurately an adversary can infer $S$ from $Z$.

We evaluate two sensitive attributes. For \emph{speaker identity}, we
train the adversary using fixed amount of enrollment audio and test
whether it can identify the speaker of unseen utterances. For
\emph{gender}, we train a binary classifier from the frozen
representations. We report balanced accuracy to account for class
imbalance and repeat each experiment across multiple randomized splits.

\vspace{-0.2cm}
\section{Evaluation}
\vspace{-0.2cm}
\subsection{Experimental Setup}

\noindent\textbf{Dataset.}
We evaluate privacy leakage using 100 two-second speech clips from
Mozilla Common Voice, comprising five speakers (three male and two
female). We use the same audio clips across all embeddings to enable a
controlled comparison.

\noindent\textbf{Audio embeddings.}
We evaluate four widely used pretrained audio representations:
CLAP, PANNs, AST, and YAMNet. For each clip, we extract the
corresponding frozen representation without fine-tuning the encoder.
Where a model produces frame-level representations, we apply temporal
mean pooling to obtain a fixed-dimensional embedding. The resulting embedding dimensions are 512 for CLAP, 2048 for PANNs, 768 for AST, and 1024 for YAMNet.

\noindent\textbf{Sensitive representations and attacks.}
For representation-level speaker leakage, we use a pretrained
ECAPA-TDNN speaker-recognition model to extract a speaker-specialized
embedding from each clip and measure its dependence with each audio
embedding. For attack-level leakage, we train an logistic regression 
probe on the frozen embeddings to predict speaker identity and gender.
We report balanced accuracy and repeat experiments across multiple
randomized splits. For speaker identification, we additionally vary the
amount of enrollment audio available to the attacker to characterize
how quickly speaker information becomes exploitable.

\noindent\textbf{Implementation details.}
For representation-level leakage, we independently
standardize the audio and ECAPA-TDNN embeddings
using \texttt{StandardScaler}. We then apply PCA
to retain 10 components from each representation,
followed by CCA with five canonical components
and a maximum of 2,000 iterations. We estimate
mutual information between corresponding canonical
components using the $k$-nearest-neighbor estimator
implemented in \texttt{mutual\_info\_regression}
($k=3$). We report the mean MI across the five
components. Statistical significance is assessed
using 100 random permutations of the paired
speaker embeddings.

\noindent\textbf{Attack configuration.}
We train a logistic regression classifier on
standardized, frozen audio embeddings using
balanced class weights, L2 regularization,$C=1.0$, and a maximum of 3,000 iterations.
For speaker identification, we randomly reserve
five clips per speaker for testing and select
one additional clip per speaker for enrollment.
We repeat this procedure 10 times using random
seeds 42-51 and report the mean and standard
deviation of balanced accuracy.

For gender inference, we use stratified random
splits with 75\% of clips for training and 25\%
for testing. We repeat the experiment over
20 randomized splits using seed 42 and report
mean balanced accuracy and standard deviation.
These splits are performed at the clip level and
may include the same speakers in training and test
sets. Therefore, the gender results measure
attribute recoverability for the evaluated speakers
rather than generalization to unseen speakers.



\subsection{Privacy Leakage Across Audio Embeddings: Results}

\noindent\textbf{Representation-level speaker leakage.}
We first examine whether general-purpose audio embeddings retain
information aligned with speaker identity. Table~\ref{tab:mi_leakage}
reports the dependence between each audio representation and
speaker-specialized ECAPA-TDNN embeddings. We observe clear differences
across representations. CLAP exhibits the highest dependence, with a
mean CCA of $0.733$ and mean MI of $0.448$, followed by PANNs
($0.325$), AST ($0.276$), and YAMNet ($0.264$) in terms of mean MI.
These results indicate that representations designed for general
acoustic understanding can retain substantial speaker-related
information, with the amount of retained information varying across
embedding models.

\begin{table}[t]
\centering
\caption{\small Representation-level speaker leakage measured by
dependence with ECAPA-TDNN speaker embeddings. Higher values indicate
greater dependence on speaker-specialized representations.}
\label{tab:mi_leakage}
\footnotesize
\setlength{\tabcolsep}{5pt}
\begin{tabular}{lccc}
\toprule
\textbf{Embedding} & \textbf{Mean CCA} & \textbf{Mean MI} & \textbf{Sum MI} \\
\midrule
CLAP   & 0.733 & 0.448 & 2.242 \\
PANNs  & 0.671 & 0.325 & 1.624 \\
AST    & 0.647 & 0.276 & 1.379 \\
YAMNet & 0.646 & 0.264 & 1.319 \\
\bottomrule
\end{tabular}
\end{table}

\begin{figure}[t]
    \centering
    \includegraphics[width=0.78\columnwidth]{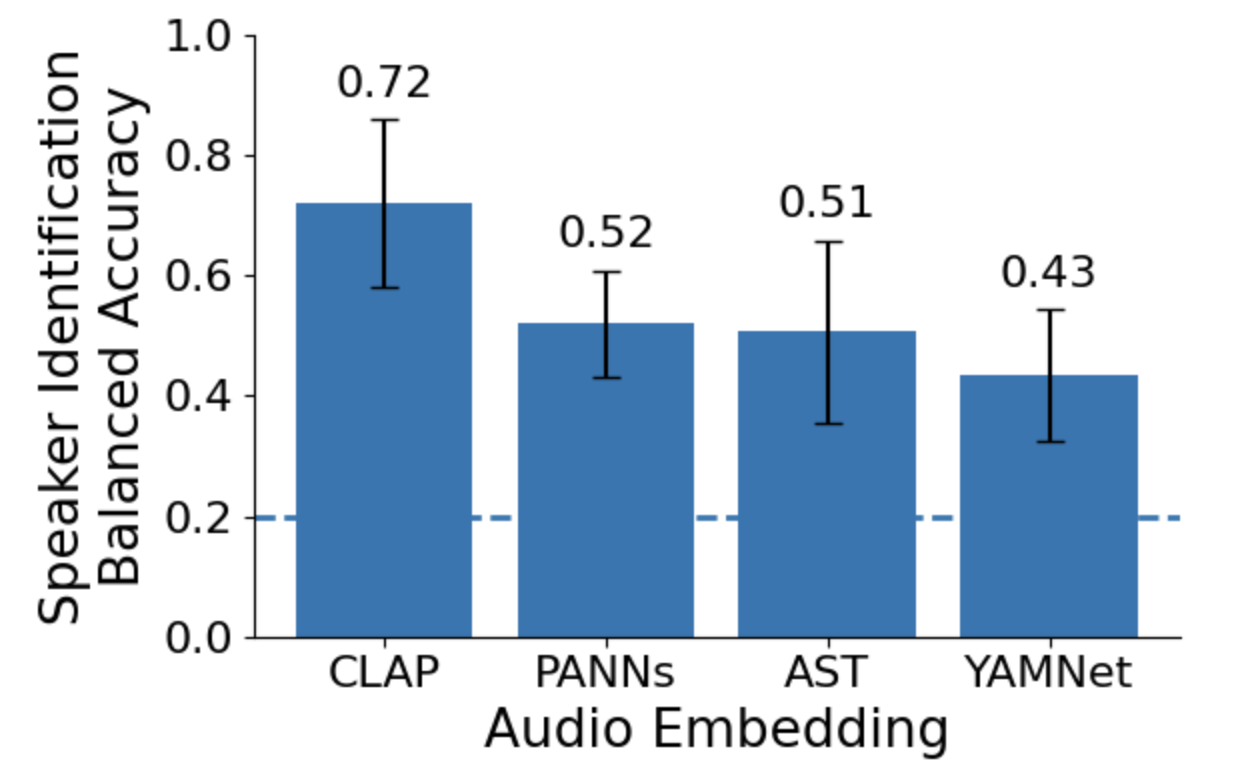}
    \caption{\small Speaker-identification attack across audio embeddings. Error bars show standard deviation
    across repeated runs; the dashed line denotes chance performance
    ($0.20$).}
    \label{fig:speaker}
\end{figure}

\noindent\textbf{Speaker-identification attack.}
We next test whether the speaker information observed at the
representation level is exploitable by an adversary. Figure~\ref{fig:speaker}
shows speaker-identification performance when the attacker is given only
one 2-s enrollment clip per speaker. All four representations exceed the
$0.20$ chance level for five-way identification. CLAP enables the
strongest attack, reaching $0.72$ balanced accuracy with only 2 seconds
of enrollment audio, followed by PANNs ($0.52$), AST ($0.51$), and
YAMNet ($0.43$). 
The representation-level and attack-level
measurements exhibit broadly consistent trends,
with CLAP showing the highest and YAMNet the
lowest speaker leakage. However, differences
between intermediate models are small relative
to the observed variability, and a larger-scale
evaluation is needed to establish whether
representation-level dependence reliably predicts
attack performance.

\begin{figure}[t]
    \centering
    \includegraphics[width=0.7\columnwidth]{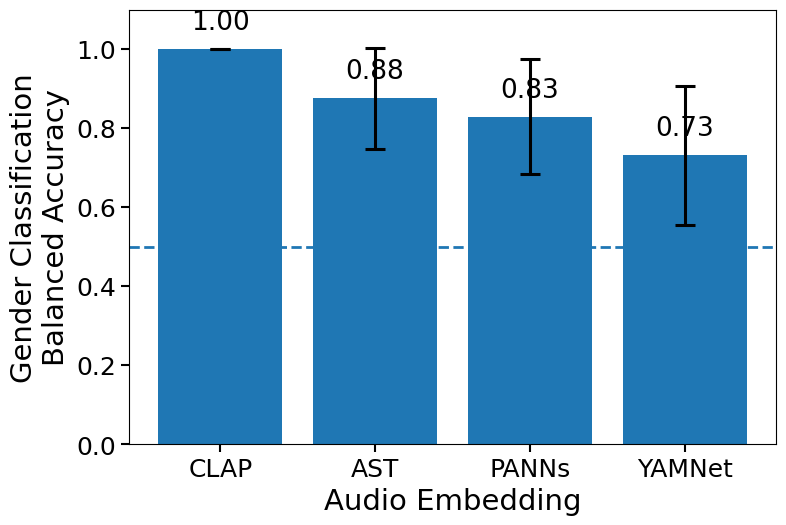}
    \caption{\small Gender inference from frozen audio embeddings.
    Error bars show standard deviation across repeated runs; the dashed
    line denotes chance performance ($0.50$).}
    \label{fig:gender}
\end{figure}

\noindent\textbf{Gender inference.}
Speaker identity is not the only sensitive information retained by
these representations. Figure~\ref{fig:gender} shows gender inference
from the same frozen embeddings. CLAP enables perfect classification
in our evaluation ($1.00$ balanced accuracy), while AST, PANNs, and
YAMNet achieve $0.88$, $0.83$, and $0.73$, respectively, all above the
$0.50$ binary chance level. Together with the speaker-identification
results, these findings show that audio embeddings intended for general
acoustic tasks can expose multiple attributes of the individuals
present in the audio, and that the degree of leakage varies
substantially across representations. 
These differences suggest that embedding
dimensionality alone does not explain privacy
leakage. Pretraining objectives, architectures,
and representation extraction strategies may
contribute to the observed variation, although
isolating their effects requires controlled
experiments.


\section{Conclusion and Future Work}

We presented an empirical framework for auditing privacy leakage in
pretrained audio embeddings using complementary representation- and
attack-level measurements. Our evaluation shows that representations
intended for general acoustic understanding can retain substantial
speaker-related information, enabling inference of speaker identity and
gender even when raw audio is unavailable. Importantly, the
representation-level and speaker-identification analyses show consistent
trends across the evaluated embeddings, with CLAP exhibiting the highest
and YAMNet the lowest speaker leakage in our experiments. These findings
highlight that intermediate audio representations should not be assumed
to be privacy-preserving and motivate privacy leakage as an explicit
criterion when selecting embeddings for audio sensing systems.
Our study is an initial audit over a small set of speakers and
representations. Future work will evaluate larger and more diverse
speaker populations, additional sensitive attributes such as speech
content, and a more audio encoders. We also plan to jointly
characterize downstream SED utility and sensitive-information leakage,
enabling analysis of the privacy-utility trade-offs of audio
representations.

\bibliography{ref}


\end{document}